\documentclass[a4paper,11pt]{article}
\pdfoutput=1 % if your are submitting a pdflatex (i.e. if you have
\usepackage{jheppub} % for details on the use of the package, please
\usepackage[T1]{fontenc} % if needed

\def\laq{~\raise 0.4ex\hbox{$<$}\kern -0.8em\lower 0.62ex\hbox{$\sim$}~}
\def\gaq{~\raise 0.4ex\hbox{$>$}\kern -0.7em\lower 0.62ex\hbox{$\sim$}~}

\def\beq{\begin{equation}}
\def\eeq{\end{equation}}
\def\bea{\begin{eqnarray}}
\def\eea{\end{eqnarray}}

\def \fp {{\dot{\phi}}}

\def \pa {\partial}
\def \ra {\rightarrow}
\def \ti {\widetilde}

\def \ls {\lambda_{\rm s}}

\def \da {\delta}

\def \ap {\alpha^{\prime}}

\def \da {\delta}

\def \fb {\overline \phi}
\def \fbp {\dot{\fb}}
\def \fbpp {\ddot{\fb}}

\title{\boldmath Non-singular pre-big bang scenarios from all-order $\ap$ corrections}

\author[a]{M. Gasperini}
\author[b]{and G. Veneziano}

\affiliation[a]{
Dipartimento di Fisica, Universit\`a di Bari, 
Via G. Amendola 173, 70126 Bari, Italy,\\
and Istituto Nazionale di Fisica Nucleare, Sezione di Bari, Italy
}
\affiliation[b]{CERN, Theory  Department, CH-1211 Geneva 23, Switzerland,\\
and Coll\`ege de France, 11 Place M. Berthelot, 75005 Paris, France\\}

\emailAdd{gasperini@ba.infn.it}
\emailAdd{gabriele.veneziano@cern.ch}

\abstract{We reformulate in Hamiltonian language the recent proposal by Hohm and Zwiebach of an action yielding the most general $O(d,d)$-symmetric string cosmology equations, at tree-level in the string-loop expansion, but to all orders in the $\ap$ expansion. This allows us to give a simple characterization of a large class of non-singular, non-perturbative, pre-big bang scenarios  smoothly interpolating between a low-energy initial accelerated (string frame) expansion and a phase of final (string and Einstein frame) decelerated expansion.  Interestingly, these solutions must  necessarily include, just around the bounce, a very short phase of (string-frame) contraction.}

\keywords{duality symmetry, string cosmology, $\ap$ corrections, bouncing solutions
 
\vskip18pt 

\noindent{\bfseries\large\sffamily{Preprints:}} BA-TH/807-23, CERN-TH-2023-066.
}

\begin{document} 
\maketitle
\flushbottom

%%%%%%%%%%%%%%%%%%%%%%%%%%%%%%%%
\section{Introduction}
\label{sec1}

Since its original proposal more than thirty years ago \cite{7} the pre-big bang (PBB) scenario (see \cite{8,9} for comprehensive reviews) has been deeply rooted on the peculiar symmetries (hereafter generically referred to  as dualities) of the tree-level string cosmology equations \cite{5,6, 1a,1c, Sen,1b,Gasperini:1993hu}.

In spite of its sound top-down motivations in string theory, and its ability to produce an observationally sensible spectrum of adiabatic curvature perturbations via the so-called curvaton mechanism (see \cite{10a,curvaton,Gasperini:2003pb} and references therein),  the PBB scenario has suffered from the lack of a convincing mechanism for avoiding the perturbative singularity that separates the pre-bounce inflationary phase from the post-bounce decelerating expansion. 

\begin{figure}[t]
\centering
\includegraphics[width=8cm]{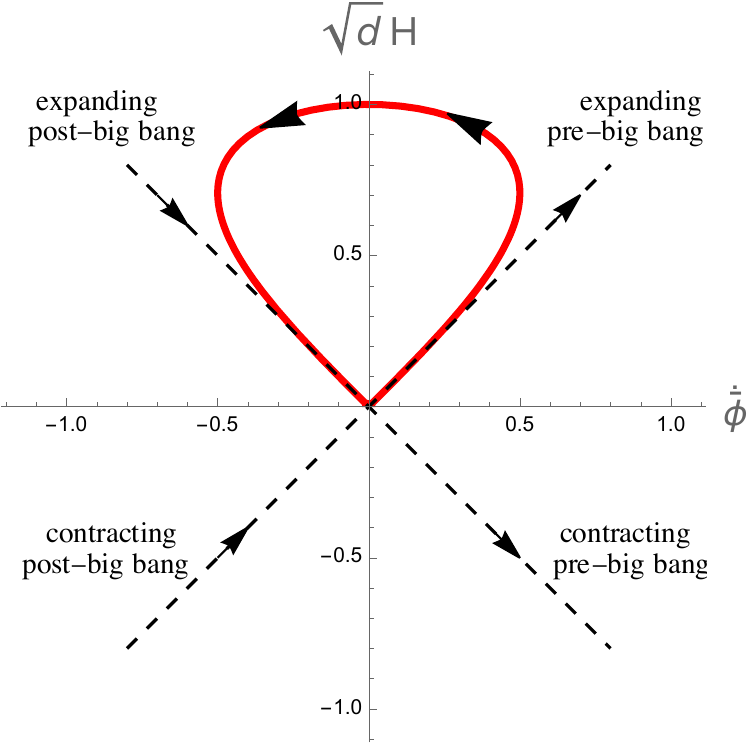}
\caption{The red curve represents the parametric plot of the solution (\ref{2}) with $d=3$. The low-energy expanding pre-big bang and post big bang branches, corresponding to the asymptotic limits $t \ra \pm \infty$ of the solution (\ref{2}), are represented by the dashed half lines $\fbp = \sqrt d H>0$ and $-\fbp = \sqrt d H>0$. Their connection implemented by the full solution (\ref{2}) describes a continuous path turning  ``anticlockwise" in the plane of the figure.}
\label{f1}
\end{figure}

So far, all promising attempts to smoothly interpolate between these two branches (see e.g. \cite{9}) are based on the addition of 
high-energy/strong-coupling corrections to the tree-level string cosmology equations, and have been mainly concentrated on searching for regular solutions with an always expanding spacetime ($H>0$) in the string frame, hence describing an ``anti-clockwise" path in the upper plane of Fig. \ref{f1}.

The path represented by the red curve  in Fig. \ref{f1}, which smoothly connects the expanding pre- and post-big bang branches ($a \sim (-t)^{-1/\sqrt{d}}$ and $a \sim (t)^{1/\sqrt{d}}$, in $d$ spatial dimensions, hence related by scale-factor duality \cite{5,6} and time reversal  $a(t) \leftrightarrow a^{-1}(-t)$ ), corresponds, in particular,  to a regular exact solution of the gravi-dilaton equations which can be explicitly parametrised, in cosmic time and in the string frame, as  \cite{10}:
\beq
a \sim \left[t+\left(1 +t^2\right)^{1/2}\right]^{1/\sqrt{d}}, ~~~~~~~~~~~~~~
\fb \sim -{1\over 2}\ln \left(1 +t^2 \right) 
\label{2}
\eeq 
(where $\fb$ denotes the generalized dilaton variable $\fb = \phi - d \ln a$ invariant under scale-factor duality transformations). 
Such a solution can be obtained by adding to the vacuum low-energy equations obtained form the string frame action the contribution of a particular (duality invariant) non-local dilaton potential \cite{10}, i.e. $V(\fb)= - V_0 e^{4\fb}$, with $V_0>0$. According to Eq. (\ref{2}) the dilaton (and thus the string coupling $g_s^2 = e^\phi$) keeps monotonically growing, so that  the appearance of additional string loop contributions may be naturally expected. Indeed, it can be shown that the above potential is only a particular case of more general class of higher-loop corrected, non local dilaton potentials  producing regular solutions (see \cite{1c}, \cite{7})  even in the presence of duality-invariant matter sources \cite{11,12}.

%%%%%%%%%%%%%%%%%%%%%%%
%%%%%%%%%%%%%%%%%%%%%

The above approach to a regular pre- to post-bounce transition, besides having possible causality problems related to non-locality, completely neglects the contribution of  
higher-curvature $\ap$ corrections. Actually, one would expect these corrections to play a crucial role for implementing a bounce  at the string scale $H \sim 1/\ls \sim 1/\sqrt{\ap}$ even in the absence of the string-loop corrections, that one will have to add eventually if the string-coupling keeps growing in time\footnote{In the solution (\ref{2}) the curvature scale at which the bounce occurs is arbitrary and, in particular, is unrelated to the string scale.}. 

If the $\ap$ corrections are included into the string frame action, and if we try to evolve the initial low-energy solution following the anti-clockwise path of Fig. \ref{f1}, it can be shown \cite{13} that we can reach continuously  a constant curvature, linear dilaton fixed-point solution located in the region $H>0$, $\fbp<0$ (at least in vacuum, and to first order in $\ap$). The price to be paid, however, is that of using higher-curvature corrected equations with explicitly broken  duality. If the symmetry is restored, as in Meissner's   first order example \cite{14}, one finds that the same fixed-point solution still exists but  cannot be smoothly connected to the low-energy pre-big bang evolution \cite{15}.

In more recent years there has been a renewed interest in analyzing and exploiting the full $O(d,d)$ symmetries of string cosmology, which generalizes scale-factor duality and which, according to \cite{Sen}, should be valid  at tree level but non-perturbatively in $\alpha'$. Explicit checks of $O(d,d)$ invariance at finite order in $\alpha'$, following the seminal paper \cite{14}, have been successfully performed  both in the conventional setup \cite{Codina1,Codina2} and in its double-field-theory reformulation \cite{Hohm1,Hohm2}.
These interesting developments, however, fall short of providing a handy formalism for dealing with the problem non-perturbatively in the $\alpha'$ expansion.

A crucial and qualitatively new step forward was recently made by Hohm and Zwiebach \cite{1} (see also \cite{21a,2,22a,3,4,25} for related studies and applications) who managed to show that $O(d,d)$ invariance, together with local field redefinitions, allow
 to recast the string cosmology equations written in the string frame in a particularly simple and manifestly duality-invariant form valid to all  orders in the $\ap$ corrections. In particular: $i)$ the so-called (duality-invariant) shifted dilaton $\fb$ appears in the very simple form that holds at lowest order; $ii)$ only first time derivatives of the metric $g_{\mu\nu}$, of the dilaton $\fb$, and of the Kalb-Ramond antisymmetric tensor $B_{\mu\nu}$ appear in the action; $iii)$ at each order in $\ap$ only a restricted set of combinations of traces of the basic $2d \times 2d$ matrix  \cite{1a} containing the spatial components $\dot{g}_{ij}$ and $\dot{B}_{ij}$  occurs.
 
In the particular case in which the Kalb-Ramond field $B_{\mu\nu}$ is set to zero (but the dilaton is non-trivial), and we specialize the string frame geometry to the case of a $(d+1)$-dimensional, homogeneous and isotropic background described by the metric $ds^2= dt^2 - a^2 (t) \da_{ij} dx^idx^j$, the $\ap$ corrected string cosmology equations for $a$ and $\phi$ depend on just one (string-theory-dependent) function of the Hubble parameter, $F(H)$, $H=\dot a/a$, which can be written as an infinite series of even powers of $H$ \cite{1}. 

The function $F(H)$ is symmetric in its argument, so that the resulting (string frame) gravi-dilaton equations are invariant under time reversal, $t \ra -t$, as well as under the scale-factor-duality subgroup \cite{5,6} of the original $O(d,d)$ symmetry, i.e. $a(t) \ra a^{-1}(t)$, $H \ra -H$, $\fb \ra \fb$. In the absence of sources, and working in the string frame, the time and space components of the gravitational equations and the equation for the dilaton field can be derived from a duality invariant action as shown in 
\cite{1}, and can be  written, respectively, as follows \cite{1,21a,2,22a,3,4}:
\beq
{\fbp}^{\,2} = F- H F', ~~~~~~~~~~~~~
\dot H F'' = \fbp F', ~~~~~~~~~~~~~
2 \fbpp = -H F' \, ,
\label{1}
\eeq
(the associated action will be explicitly recalled in Sect. \ref{sec3} for possible generalizations.)
Here the dot denotes cosmic-time derivative, and a prime (following the notations of \cite{1}) denotes the derivative with respect to the Hubble parameter, $F' = dF/dH$. It may be useful to note that, if $\fbp \not= 0$, the dilaton equation for $ \fbpp$ is a direct consequence of the first two equations.
Note also that the function $F$, to zeroth-order in the $\ap$ expansion,  simply reduces to $F= -d H^2$, and one recovers from Eqs. (\ref{1}) the low-energy string cosmology equations with the well-known solutions describing the phase of  ``dilaton driven" inflation and its dual  \cite{7,8,9} , namely $a \sim (\mp t)^{\mp 1/\sqrt{d}}$, $\fb = - \ln (\mp t)$, $\fbp = \pm \sqrt{d} H$ , represented by the dashed bisecting lines of Fig. \ref{f1}. 

In this short note we present and discuss a class of alternative bouncing scenarios, based on regular solutions of the equations (\ref{1}), invariant under scale factor duality, and implementing the transition from the low-energy expanding pre-big bang and post-big bang solutions by turning ``clockwise" in the $\{\fbp, \sqrt d H\}$ plane. We will give quite generic sufficiency conditions for such solutions to exist. We will also argue that,
 if we assume the duality symmetry to be preserved at all orders in $\alpha'$, and the background evolution (in vacuum) is thus described by Eqs. (\ref{1}), it is impossible to obtain solutions describing an anti-clockwise close-loop trajectory like the one of Fig. \ref{f1}.
 
 In Sect. \ref{sec2} we will restrict the discussion to the isotropic case for which Eqs. (\ref{1}) apply. We will also discuss the form and regularity of the solutions in the Einstein frame. Then, in Sect. \ref{sec3}, we present a strategy for extending our considerations  to a generic anisotropic Bianchi I cosmology.  The above-mentioned argument for the non existence of anti-clockwise solutions will be detailed in  Appendix \ref{app}.

%%%%%%%%%%%%%%%%%%%%%%

\section{Regular isotropic bouncing solutions}
\label{sec2}

\begin{figure}[t]
\centering
\includegraphics[width=8cm]{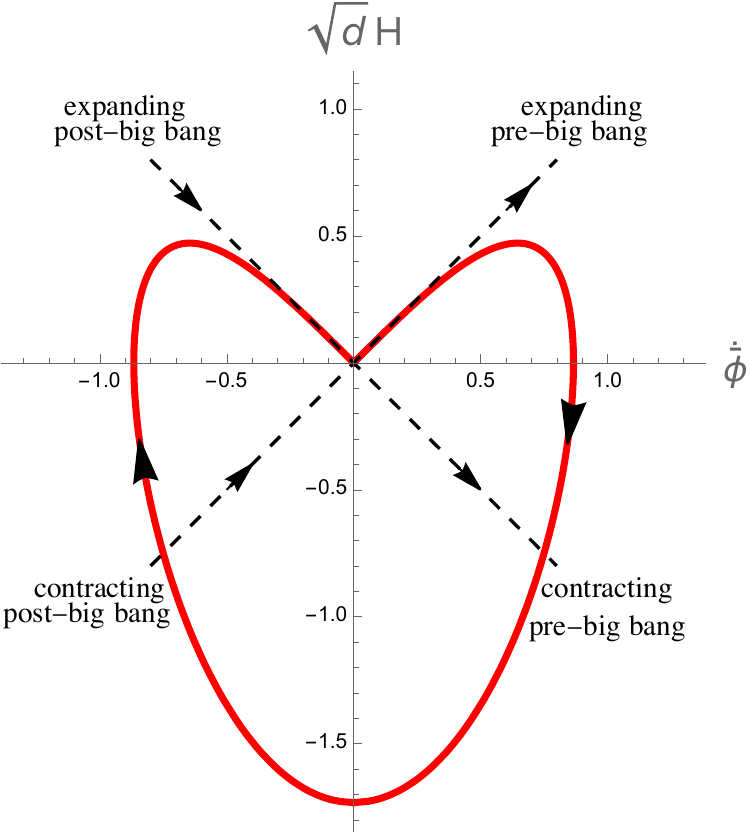}
\caption{Parametric plot of the regular bouncing solution of Eq. (\ref{3}), with $\ap=2$ and $d=3$. It describes a continuous, heart-like path turning clockwise in the plane of the figure.}
\label{f2}
\end{figure}

The work presented in this paper was prompted  by a class of regular bouncing solutions of Eqs. (\ref{1}) recently presented in \cite{2} as a result of some ``trial and error" procedure. For the bosonic string model the solution can be parametrized directly in cosmic time as follows \cite{2}:
\beq
H(t)=\sqrt{2}{2dt^2- \ap\over (2dt^2 + \ap)^{3/2}}, ~~~~~~~~~~~~
\fb(t)=\ln \left[{1\over 2} \left(\ap \over d (2dt^2 + \ap)\right)^{1/2}
\right].
\label{3}
\eeq
The corresponding parametric plot, shown in Fig. \ref{f2}, describes a ``heart-shaped" curve travelled clockwise as time goes on. It is not clear, from the discussion in \cite{2}, which exact choice of $F(H)$ leads to the solution (\ref{3}), although the authors make some consistency perturbative checks. It is also unclear whether a special fine-tuning of the initial data and/or of $F$ are needed in order to avoid the singularity.

%%%%%%%%%%%%%%%%%%%%%%%
%%%%%%%%%%%%%%%%%%%%%

 Below, we will clarify this issue by spelling out  sufficient (and probably necessary, see Appendix \ref{app}) conditions for a regular ``clockwise bounce" to occur. We find,
amazingly, that the clockwise alternative turns out to be much easier to achieve than its anti-clockwise counterpart, and that the conditions for this to happen are rather generic and easy to implement restrictions on the arbitrary function $F(H)$. In such a context, 
it looks however necessary for $F(H)$ to have an $\alpha'$ expansion with a finite radius of convergence, due to the presence of branch-points in the complex-$H$ plane.  We think that the possible existence of singularities at values of $H$ of order $1/\sqrt{\ap}$ is to be expected, in general, since the effective action describing the dynamics of the massless fields comes from integrating out the string's massive degrees of freedom, and such an effective action -- like in any quantum field theory -- will cease to be analytic when the energy scale (here $H$) becomes comparable to the mass of the lightest massive mode. This means, in other words, that our regular-bounce solutions are truly non-perturbative from the point of view of the $\ap$ expansion (unlike some other cases recently considered in \cite{25}). 

We stress again that, in order to get a non-singular bounce, there is no need to invoke loop corrections, to break scale-factor duality, and/or to add non local dilaton potentials. All we need is to assume string theory to be kind enough to lead to an  $F(H)$ satisfying two properties that we shall now spell out for this simplest (FLRW) geometry (see Sect. \ref{sec3} for an extension to anisotropic backgrounds).

\begin{figure}[t]
\centering
\includegraphics[width=8cm]{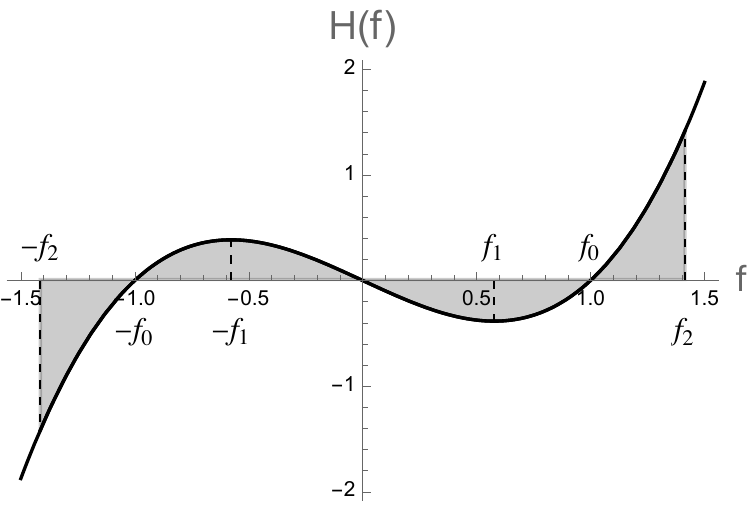}
\caption{Qualitative behaviour of a simple function $H(f)$ satisfying the two conditions (\ref{4}) and (\ref{5}). The plotted curve $H(f)= -f+f^3$ is   characterised by  $f_0=1$, $f_1=1/\sqrt{3}$, $f_2= \sqrt{2}$, and physically corresponds to the particular solutions presented in \cite{2} for appropriate values of the parameters.}
\label{f3}
\end{figure}

It is convenient to start by defining the function $f(H)\equiv  F'(H)$.  Because of the symmetry properties of $F$ we have $f(H)=- f(-H)$. At zeroth-order in $\ap$ we know that $f(H)= -2 d H$, so that, in this limit, we can easily invert the function $f(H)$ to obtain $H(f)= -f/2d$ for $\ap \ra 0$. 
In order to characterise the class of functions implementing a  ``clockwise" regular bounce let us consider the inverse function $H= H(f)$ at the non-perturbative level in $\ap$. By definition, it satisfies the symmetry property $H(f)=-H(-f)$. 
Let us concentrate on the class of functions $H(f)$ satisfying also the two following conditions (see Fig. \ref{f3}):
\begin{enumerate}
\item{}
 $H(f)$ has a zero\footnote{For the bosonic and heterotic string models, the sign of the first-order $\ap$ term appearing in $F$ is consistent with this assumed condition  \cite{1}.}
 at $f= \pm f_0 \not=0$. Since $|H|$ is initially growing this is true provided there is some point $f_1$, with $|f_1| <|f_0|$, where the function $H(f)$ has an extremum, i.e.
 \beq 
 \left(\pa H/\pa f\right)_{f_1} =0, ~~~~~~~~~~ |f_1| <|f_0|.
 \label{4}
 \eeq
 \item{}
For $f>f_0$ and $f<-f_0$ the function $H(f)$ will typically change sign, and the second condition is that there is a point $f_2>f_0$ (and by symmetry one at $f_2 < - f_0$) such that
\beq
\int_0^{f_2} dfH(f) = 0 =  \int_{-f_2}^0 df H(f) .
\label{5}
\eeq
\end{enumerate}
Such a behaviour of the function $H(f)$ is qualitatively illustrated in Fig. \ref{f3}, where the shaded areas from $0$ to $f_0$ and from $f_0$ to $f_2$ are exactly the same but with opposite sign, so that the integral (\ref{5}) is vanishing. 

%%%%%%%%%%%%%%%%%%%%%%%
%%%%%%%%%%%%%%%%%%%%%

It is clear that the two above conditions in no way imply a fine-tuning of the function $H(f)$. We now claim that the class of functions satisfying such conditions implement a regular bouncing pre-big bang scenario exactly satisfying the equations (\ref{1}), and of the type illustrated in Fig. \ref{f2}, where:
\begin{itemize}
\item{} 
$H(-f_1)$ is the positive maximum of $H$ (the two upper ``bulges"  of the ``heart" of Fig. \ref{f2});
\item{}
$-f_0$ corresponds to the point at which $H$ turns negative (the intersection  of the ``heart" with the positive axes  $\fbp$ in Fig. \ref{f2}), with $\fbp$ reaching its own maximum; and 
\item{}
$-f_2$ is where $H$ reaches its negative minimum (the bottom ``bulge" of the ``heart" of Fig. \ref{f2}).
\end{itemize}
At the negative minimum of $H$ we also have $\fbp=0$ (see Fig. \ref{f2}), so that $\fbp$ turns negative at $f=-f_2$, and it always stays negative at later times when $f$ goes back to the origin following the previous route backwards\footnote{Actually, in order to avoid singularities in the solution, the ``inverted route" must include some change in the branch to be used for the function $F(H)$. This is where the presence of branch points in $F(H)$ plays a crucial role. We are grateful to Barton Zwiebach for having emphasized this point with us in a private communication.}. 

Note that, by differentiating with respect to $H$ the general function $F-HF'$ appearing in the equations (\ref{1}), one easily obtains
\beq
F-H F'= -\int_0^H \ti H f'(\ti H) d\ti H = - \int_0^{f(H)} H(\ti f) d \ti f.
\label{6}
\eeq
Hence, for $f \ra - f_2$, the condition (\ref{5}) implies $F-HF' \ra 0$, in agreement with the first of the equations (\ref{1}) and with the fact that, at $f=-f_2$, we also have $\fbp=0$. Even more important is the fact that, by using Eqs. (\ref{1}) and (\ref{6}), we can express $\fbp$ in terms of $f$ as
\beq
\fbp (f) = \pm \left[ -\int_0^f H(\ti f) d \ti f  \right]^{1/2}.
\label{7}
\eeq
Once $H(f)$ is given then also $\fbp(f)$ is known, and we can use the variable $f$ to perform parametric plots of the solution in the plane $\{\fbp(f), \sqrt{d} H(f)\}$, to be compared with the one Fig. \ref{f2}. 

As an application of the above results, let us give two simple examples satisfying our criteria (\ref{4}) and (\ref{5}) for regular exact solutions of all-orders equations (\ref{1}), and describing clockwise close-loop trajectories in the plane of Fig. \ref{f2}. 

The first example is our way to reconstruct the bouncing solution of \cite{2} starting from a suitable (and simple) form\footnote{Instead, the explicit from of $F(H)$, the solution of a cubic equation endowed with several branch points, would have been extremely difficult to guess.} of 
 $H(f)$. If we satisfy the conditions (\ref{4}) and (\ref{5}) by choosing
\beq
H(f)= -{f\over 2d} + \ap \left(f\over 2d\right)^3,
\label{8}
\eeq
we obtain, from Eq. (\ref{7}),
\beq
\fbp(f) = \pm {f\over 2d} \left( d- {\ap\over 8} f^2\right)^{1/2},
\label{9}
\eeq
and we can easily check that the parametric plot of this solution exactly coincides with the plot shown in Fig. \ref{f2} for the solution (\ref{3})  given in cosmic time, as in the original parametrisation \cite{2}. 

Let us now consider the following (new) example corresponding to
\beq
H(f)={ -{f/2d} + (\ap/2)\left(f/ 2d\right)^3\over 1+ (\ap/2)(f/2d)^2}.
\label{10}
\eeq
Again, the conditions (\ref{4}) and (\ref{5}) are satisfied, and we can use Eq. (\ref{7}) to compute $\fbp(f)$ and the associated parametric plot for this solution. However, let us first illustrate the interesting behavior of the Hubble parameter around the bounce, by plotting $H$ and $\fbp$ as a function of the cosmic time  $t$. 

In order to obtain $H(t)$ we need $f(t)$. By definition $\dot f \equiv f' \dot H$, so that  the second of the equations (\ref{1}) gives us $\dot f = f \fbp$. On the other hand, we also have $\dot H^{-1} \equiv dt/dH= (dt/df)(df/dH)\equiv f'/f \fbp$. 
Hence, by using Eq. (\ref{7}) for $\fbp$, we get
\beq
dt={f'dH\over f \fbp}= \pm {df\over f}  \left[ -\int_0^f H(\ti f) d \ti f  \right]^{-1/2}.
\label{11}
\eeq
By integrating and inverting this equation, for any given $H(f)$, one can then obtain $f(t)$ and the corresponding time behaviour of $H(f(t))$ and $\fbp(f(t))$.

%%%%%%%%%%%%%%%%%%%%%%%
\begin{figure}[t]
\centering
\includegraphics[width=6.5cm]{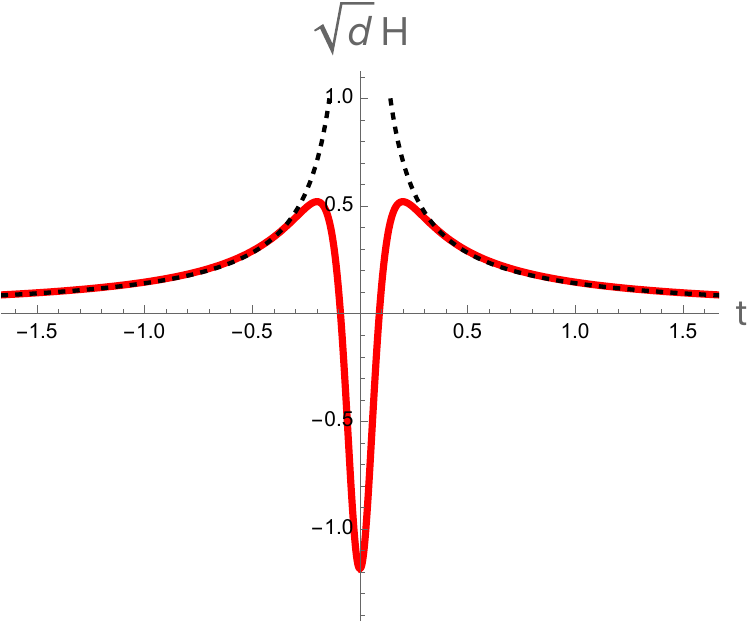}~~~~~~~~~~
\includegraphics[width=6.5cm]{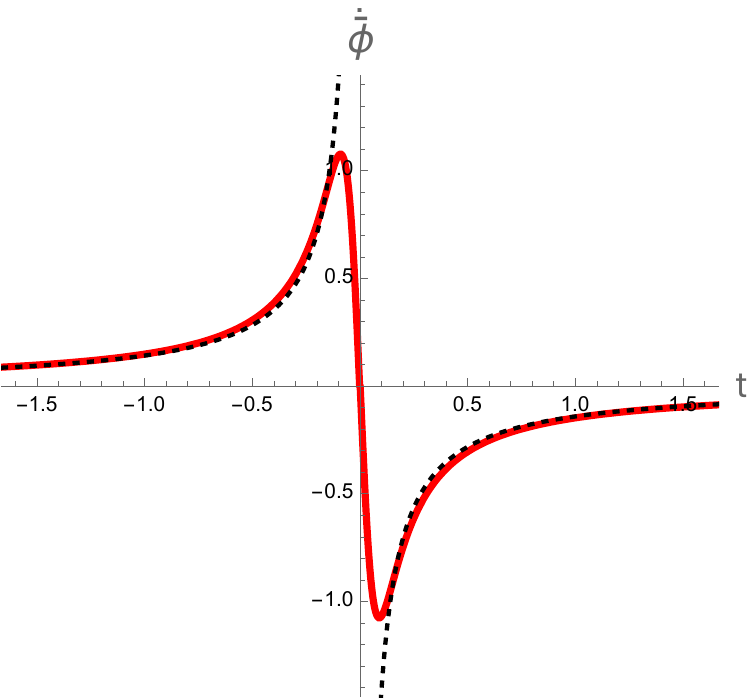}
\caption{The time behaviour of the solution (\ref{10}),  (\ref{12}) is represented by the solid red curves, plotted for $d=3$ and $\ap=2$ (for graphical reasons, we have expressed time in units of $5 \sqrt{\ap}$). The dashed black curves describe the behaviour of the (singular) low-energy pre- and post-big bang branches, corresponding to the asymptotic limits $t \ra \mp \infty$.
The smooth transition between the initial accelerated to the final decelerated expansion is triggered by a high-curvature phase of accelerated/decelerated contraction.}
\label{f4}
\end{figure}
%%%%%%%%%%%%%%%%%%%%%

For the solution (\ref{10}) such a procedure can be easily applied numerically, and the results is shown in Fig. \ref{f4} (where we have set $d=3$ and $\ap=2$). The oscillation of sign of $H$, localised in the short bouncing region should be underlined: a similar effect is also present in the solution reported in \cite{2}, and it is typical of a scenario where the smooth connection between the pre- and post-big bang branches is implemented through a heart-like curve, turning clockwise in the plane $\{\fbp, \sqrt d H\}$. 

\begin{figure}[t]
\centering
\includegraphics[width=6.5cm]{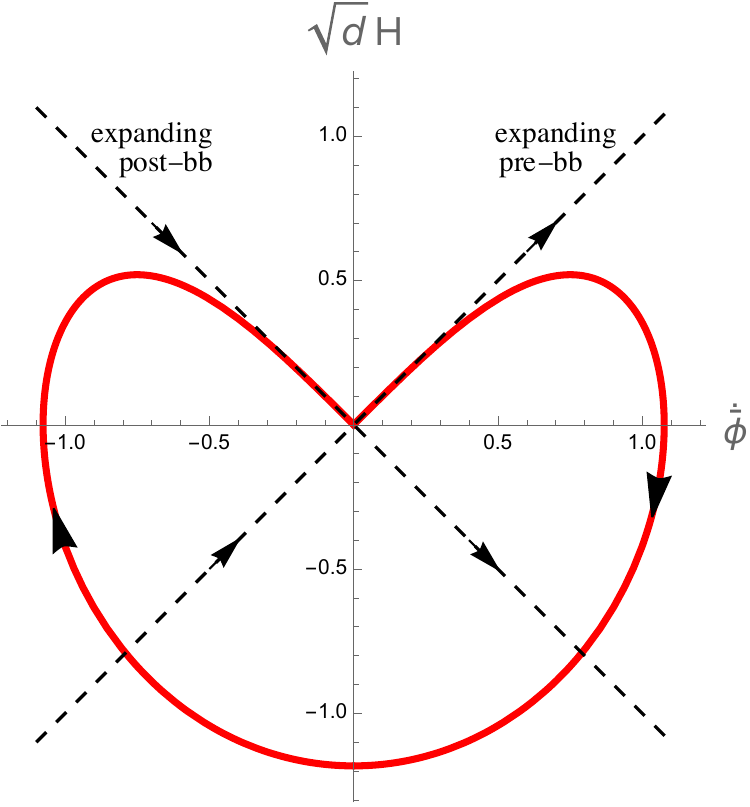}
\caption{Parametric plot of the new solution (\ref{10}), (\ref{12}) (for $d=3$, in units $\ap=2$). We stress that the direction of the arrow along the curve cannot be reversed.}
\label{f5}
\end{figure}

It seems appropriate to confirm this point by giving also the parametric plot, in the above plane, of the solution $H(f)$ corresponding to Eq. (\ref{10}). The numerical integration of Eq. (\ref{7}) gives $\fbp (f)$, which can be written (again, for $d=3$ and $\ap=2$) as
\beq
\fbp(f) = \pm \left[6 \ln \left(1+\frac{f^2}{36} \right) - {f^2\over 12} \right]^{1/2}. 
\label{12}
\eeq
The parametric plot of the solution (\ref{10}), (\ref{12}) then gives  the red curve of Fig. \ref{f5}. We can easily check that the two upper bulges and the bottom bulge of the plotted heart exactly correspond, respectively, to the positive maximum and negative minimum of $H$ appearing in Fig. \ref{f4}.

%%%%%%%%%%%%%%%%%%%%%%%
%%%%%%%%%%%%%%%%%%%%%

Let us finally briefly discuss a few kinematic properties of this regular and isotropic bouncing scenario. 
First of all we notice that we have considered solutions (to all orders in $\ap$) of the string cosmology equations (\ref{1}) written in the so-called string-frame \cite{9}, where the initial, asymptotic form (at $t \ra -\infty$) of the solution describes an accelerated pre-big bang expansion (the bisecting line of the upper-right quadrant of Fig. \ref{f5}). What happens if we describe this scenario in the context of the more conventional Einstein-frame geometry, where the gravi-dilaton kinetic action is canonically normalised?

\begin{figure}[t]
\centering
\includegraphics[width=8cm]{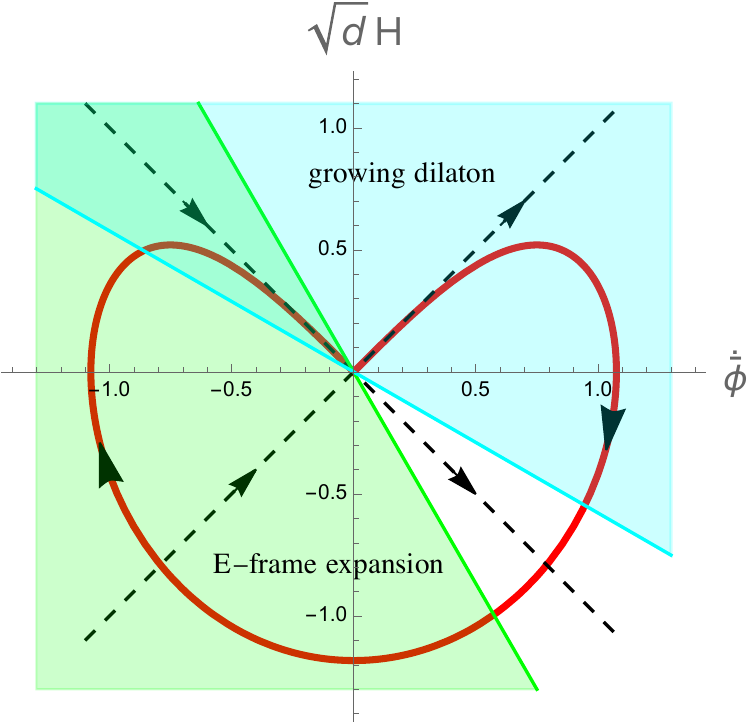}
\caption{The green shaded region, satisfying the condition $H<-\fbp$, is the allowed region for solutions describing an expanding metric in the E-frame. 
The sky-blue shaded region, satisfying the condition $H>-\fbp/d$, is the allowed region for solutions describing a growing string coupling. The superposition of the two allowed regions, in the upper-left quadrant of the figure, contains the final, expanding post-big bang regime of the bouncing solutions presented in this paper.}
\label{f6}
\end{figure}

It is well known (see e.g. \cite{9}) that for a homogeneous and isotropic geometry the E-frame and S-frame scale factors, $a_E$ and $a$, are related by the conformal transformations $a_E \sim a e^{-\phi/(d-1)}$. It follows that $H_E \sim H - \fp/(d-1)$, where $\fp= \fbp +dH$. Hence, an expanding geometry in the E-frame, $H_E>0$, is characterised by the condition $H<-\fbp$, which is satisfied in the green shaded region on the left of the green line plotted in Fig. \ref{f6}. Thus, in the E-frame, we have an initial accelerated contraction (i.e. with negative first and second time derivatives of the scale factor), which turns  into expansion just before the curvature bounce, and keeps expanding up to the final, decelerated, low energy regime (i.e. with positive first and negative second time derivatives of the scale factor). Note that the bounce of the scale factor (turning contraction into expansion or viceversa) is not necessarily exactly simultaneous with the bounce of the  curvature (turning a phase of growing into decreasing spacetime curvature), even if they both occur at the string scale (as also shown by the time behaviour of the Hubble parameter in Fig. \ref{f4}).

%%%%%%%%%%%%%%%%%%%%%%%
%%%%%%%%%%%%%%%%%%%%%

Second, we know that the dilaton is growing ($\fp>0$) in the initial, asymptotic solution: what about the dilaton in the final regime? From the definition of $\fbp$, the condition of a growing dilaton is given by $H>-\fbp/d$, which is satisfied in the sky-blue shaded region on the right of the sky-blue line plotted in Fig. {\ref{f6} (including the final expanding post-big bang regime).

%%%%%%%%%%%%%%%%%%%%%%%
\begin{figure}[b]
\centering
\includegraphics[width=8cm]{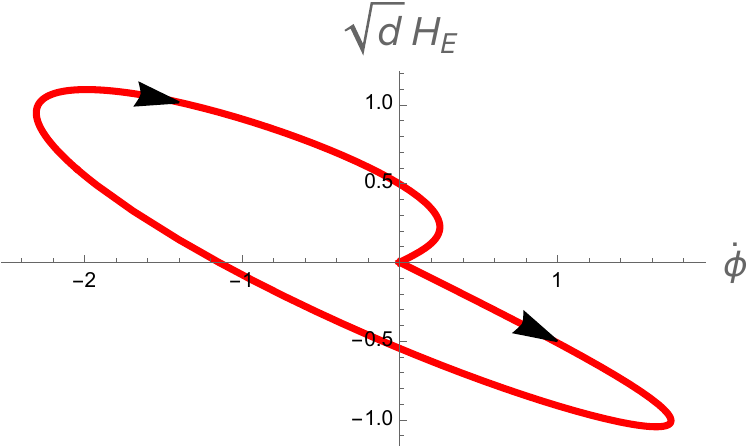}
\caption{The ``deformed-heart" curve representing the parametric plot of the solution (\ref{10}), (\ref{12}) (for $d=3$, in units $\ap=2$), in the plane spanned by the  kinetic energy of the ``physical" dilaton field $\phi$ and by the Hubble parameter $H_E$ of the Einstein-frame geometry.}
\label{f6a}
\end{figure}
%%%%%%%%%%%%%%%%%%%%%

The above properties of the Einstein-frame geometry and of the dilaton dynamics can be illustrated even more clearly by performing the parametric plot of the above solution in the plane spanned by the variables $\{\dot \phi, \sqrt{d} H_E\}$, and represented by the ``deformed-heart" curve of Fig. \ref{f6a}. Again, the trajectory of the curve is turning clockwise in the plane of the figure, but there is only one point with vanishing $H_E$, where the initial contraction $H_E<0$ turns into an expansion. Also, the time evolution of $\dot \phi$ is not 
left-right symmetric like that of the duality invariant-variable $\fbp$ but, remarkably, its behaviour is qualitatively the same both in the string and Einstein frame (unlike that of $\fbp$). 
Let us notice, finally, that having a bounce of the curvature in the Einstein frame implies an {\it effective} violation of some energy conditions if one brings the higher-derivative terms of the string cosmology equations on the r.h.s. of the standard Einstein equations,  interpreting them as some effective matter contribution. But, actually, the singularity regularisation, and the smooth transition from pre- to post-big bang, is not due to exotic sources but to the modified (with respect to Einstein's) gravitational dynamics produced by the $\ap$ corrections.

Therefore, in this scenario, the initially growing dilaton (i.e. string coupling) starts decreasing shortly before and during the bounce, but it increases again in the final regime after the bounce. This last property is a welcome one in our context, because it naturally leads to a final strong coupling regime where the string-loop corrections are expected (see e.g. \cite{9}), and actually needed in order to induce particle production (reheating) effects, to stabilise the string coupling, and to drive the cosmological background to the standard post-bounce evolution scenario. In this respect, the results presented here should be regarded only as a preliminary step towards a more complete and realistic description of the post-bounce Universe, addressing phenomenological consequence and observational constraints.

%%%%%%%%%%%%%%%%%%%%%%%%%%%%%%%%%%%%%

\section{A  non-perturbative strategy for finding regular bouncing solutions}
\label{sec3}

In this Section we will present a more general approach to the problem of finding regular bouncing solution using $\ap$ corrections to all orders and even beyond the radius of convergence of its Taylor series. We will illustrate the procedure in the case of a generic anisotropic Bianchi I, higher-dimensional background, characterised by the line element 
$ds^2= dt^2 - \sum_i a_i^2 (t)  dx^idx^i $, where $i= 1, \dots d$, 
to which we add, as before, a time-dependent dilaton $ \phi(t) $ 
(see also \cite{25,26,rost,27} for a discussion of anisotropic backgrounds).

In this case the basic outcome  of Hohm and Zwiebach's analysis \cite{1}  can be simply formulated in terms of an action $S$ depending on the (log of the) scale factors, $\beta_i \equiv \ln a_i$, on the shifted dilaton $\bar{\phi}$, and on the lapse function $N$ as follows:
\beq
S= \int dt N e^{-\bar{\phi} } \left [ N^{-2} \fbp^2 + F(N^{-1} \dot{\beta}_i) \right]~~~ ;~~~~~~
~~~ F = - N^{-2} \sum_i \dot{\beta}_i^2 + \cdots 
\label{211}
\eeq
where the dots denote higher order $\ap$ corrections. 
Defining as usual $H_i = \dot{\beta}_i$, the duality group now becomes the $Z_2^d$ group, corresponding to the transformations:
\beq
 H_i \rightarrow  -H_i ~~ ; ~~~~~~~~~~~~~~
 \bar{\phi} \equiv \phi - \sum_i \ln a_i  \rightarrow \bar{\phi}, 
\label{212}
\eeq
forcing the function $F$ to satisfy $F(H_i)= F(-H_i)$. The generalisation of the field equations (\ref{1}) is easily obtained by varying the action w.r.t. its variables and reads:
\beq
{\fbp}^{\,2} = F- \sum_i H_i f_i ~;~~~~~~~~~~~~
 f_{ij}\dot{H}_j \equiv \dot{f_i} = \fbp f_i ~;~~~~~~~~  2 \fbpp = - \sum_i H_i f_i \, ,
\label{213}
\eeq
where we have defined $f_i = {\partial F}/{\partial H_i}$  and $f_{ij} = {\partial f_i/}{\partial H_j} = f_{ji}$.

Our proposal for searching for regular bouncing solution consists of characterizing, rather than the ``Lagrangian" $F$, the ``Hamiltonian" $h$ connected to it by a standard Legendre transform. Denoting the momenta conjugate to $\beta_i$ by $\pi_i$ we have (after eventually setting $N=1$):
\beq
\pi_i = \frac{\pa F}{\pa \dot{\beta}_i } = - 2 H_i + \dots  ~;
~~~~~~~~~~ h(\pi_i) = \sum_i  \pi_i H_i - F(H_i) =  -\frac14  \sum_i  \pi_i^2 + \dots
\label{214}
\eeq
As usual, the Hamiltonian $h$ satisfies the exact equation ${\pa h}/{\pa \pi_i} = H_i$ which basically inverts the relation between the $f_i$ and the $H_i$. We also note that $h(\pi_i)$ is nothing but the function $g(H_i)$ of \cite{1}, 
i.e. $g(H_i)= \sum_k f_kH_k -F$, simply expressed in terms of the conjugate variables $\pi_i \equiv f_i$.

 Our strategy consists in spelling out conditions on $h(\pi_i)$ (rather than on $F(H_i)$) insuring that a regular bouncing solution will exist. As it turns out, these conditions are easily satisfied by very simple Hamiltonians. These, however, correspond to complicated Lagrangians exhibiting branch point singularities and therefore having Taylor expansions with finite radius of convergence. 
Before illustrating this procedure  in a specific example, it is convenient to use appropriately rescaled momenta $z_i \equiv - {\pi_i}/{2}$ so that, at lowest order in $\alpha'$, 
\beq
H_i(z_j) = z_i + \dots ~;~~~~~~~~~~~ h(z_j) = -  \sum_i z_i^2 + \dots  ;~~~~~~~~~~~~~  \frac{\pa h(z_j)}{\pa z_i} = - 2 H_i \, .
\label{213a}
\eeq

Let us now imagine to be given the function $h(z_i)$ and let us rewrite the three equations (\ref{213}) in terms of it.
We easily obtain:
\beq
{\fbp}^{\,2} = - h(z_i)  ~;~~~~~~~~~~~~
 \dot{H}_k = - \frac12 \frac{\pa^2 h}{\pa z_k \pa z_l } z_l \fbp ~;~~~~~~~~~~~~   2 \fbpp =  - \sum_i  z_i \frac{\pa h}{\pa z_i} \, .
\label{215}
\eeq
We can also obtain a simple evolution equation for $z_i$ since, by definition, $z_i= - \pi_i/2= -f_i/2$ so that, by using the second of Eqs. (\ref{213}) for $\dot f_i$, we find $\dot z_i= z_i \fbp$. 
In order to present a parametric plot of the various $H_i$ in terms of $\fbp$ it is then sufficient to use the Hamilton equation $H_i(z)= \pa h/\pa \pi_i= -(1/2) \pa h/ \pa z_i$, and to express $\fbp$ in terms of the $z_i$ using the first of Eqns. (\ref{215}).

A simple class of regular anisotropic bounces can now be obtained by generalizing the ansatz (\ref{8}) as follows\footnote{This is the simplest example in which the different scale factors interact with each other. We are aware of the fact that, according to \cite{1},  ``double-trace" terms, such as the one multiplying $c_3$, only occur at higher orders in $\alpha'$.}:
\beq
h( z_i) =  - \sum_i z_i^2  + \ap \frac{c_2}{8}  \sum_i z_i^4 + \ap  \frac{c_3}{8}  \left(\sum_i z_i^2\right)^2 ,
\eeq 
from which
\beq
 H_i   \equiv -{1\over 2} {\pa h\over \pa z_i}
=  z_i  - \ap \frac{c_2}{4}  z_i^3 -  \ap \frac{c_3}{4}  z_i \sum_j z_j^2 \, ,
\label{216}
\eeq
where, in analogy with what we did in the isotropic case, we take $c_2, c_3 >0$ in order to generalise our conditions for a regular bounce.
In Fig. \ref{f7} we show an example of anisotropic, yet regular, bounce,
for a $d+n$ dimensional space where $d$ dimensions are expanding with scale factor $a_1$, and $n$ dimensions are expanding with scale factor $a_2$. We have numerically integrated the evolution equations $\dot z_i= z_i \fbp$} by imposing initial conditions at $ H_i \ra 0$ according to the low-energy pre-big bang solution (which satisfies $d H_1^2 + n H_2^2 = \fbp^2$), and we have chosen, initially, $H_1 \not= H_2$.

%%%%%%%%%%%%%%%%%%%%%%%%
\begin{figure}[t]
\centering
\includegraphics[width=8cm]{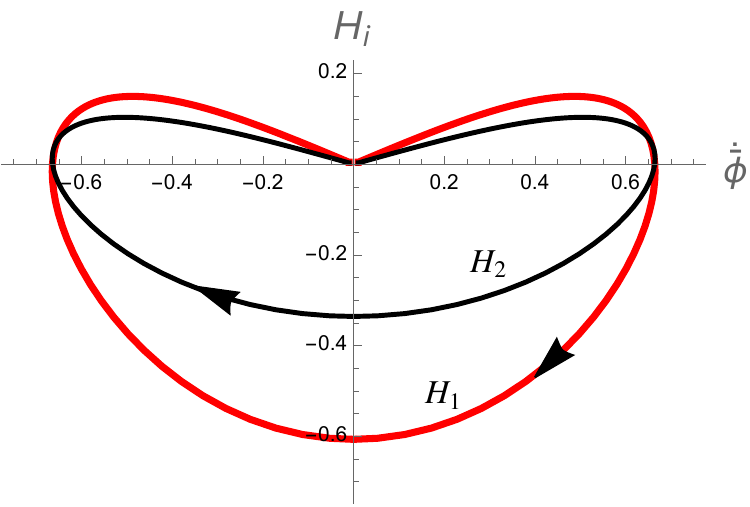}
\caption{Parametric plot of the two Hubble parameters corresponding to the case of Eq. (\ref{216}) in which, out of a total of nine $H_i$, $d=3$ are equal to $H_1$ (red curve) and $n=6$ are equal to $H_2$ (black curve). We use units in which $\ap=2$ as well as $c_2 = c_3 =2$.  We took, as initial conditions,  ${H_1}/{\fbp} = \sqrt{(1 + n  \epsilon) /(d+n)}$, ${H_2}/{\fbp} =\sqrt{ (1 - d  \epsilon)/(d+n)}$, with $\epsilon = 0.1$ .}
\label{f7}
\end{figure} 
%%%%%%%%%%%%%%%%%%%%%%%%

We may note that the different $H_i$ ``bounce" (i.e. reach their minimal negative value) at the same time, when $\fbp=0$. However,  the positive maximal value of the $H_i$, as well as the bounces of the different scale factors (where $H_i=0$) are not reached simultaneously along the different spatial dimensions. The first of these  two properties of the solution (\ref{216}) is clearly displayed by the plot of Fig. \ref{f7} while the second is not (because the actual numbers are very close), but  can be easily deduced by the general equations (\ref{213}). 

Also, it seems that there is no isotropisation mechanism induced by the $\alpha'$ corrections (at least in the case of this simple example). This last effect is even more evident if we assume, for instance, that initially $d$ dimensions are expanding while the other $n$ dimensions are contracting (or vice versa), because such a different kinematic behaviour will  characterise also the final asymptotic configuration after the bounce. Indeed, our regular bounce connects two Kasner-type cosmologies -- actually, their  generalisation including a massless scalar field --  that are related by the simplest subgroup of $O(d,d)$, the scale-factor duality transformation that flips the sign of all the Kasner exponents. Amusingly, this is just what has been recently selected as  a possible consistent matching \cite{LLV} between data across the two sides of a singular hypersurface. 
We plan to come back to a more complete study of the anisotropic situation (with both expanding and contracting dimensions) in the near future.

%%%%%%%%%%%%%%%%%%%%%%%%
\begin{figure}[t]
\centering
\includegraphics[width=8cm]{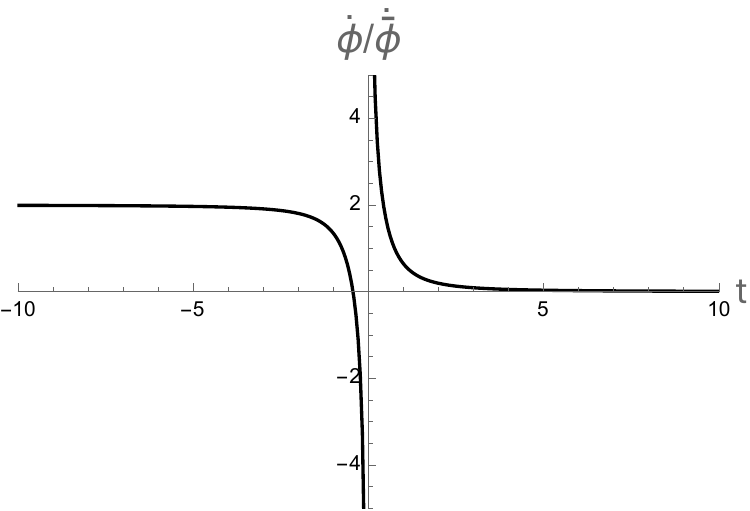}
\caption{Time evolution of the dilaton's  growth rate according to the solution  (\ref{216}), for the particular anisotropic initial configuration described in the text. Quite late after the bounce epoch the dilaton tends to become time-independent (the apparent divergence at $t=0$ is due to the vanishing of $\fbp$ at the bounce, see e.g. Fig. \ref{f7}).}
\label{f7a}
\end{figure} 
%%%%%%%%%%%%%%%%%%%%%%%%

Here we shall only mention that in such an anisotropic context it is possible to interpolate smoothly between a  rapidly varying dilaton in the initial phase and an essentially constant dilaton  in the far future\footnote{We thank Robert Brandenberger for bringing up the question of  dilaton stabilisation, which motivated this last comment.} (as illustrated in Fig. \ref{f7a}). For this to happen one just needs to impose  that the final Hubble parameters satisfy the condition $\dot \phi = \fbp + \sum H_i \to 0$ for $t \ra +\infty$, which, in view of the flip of sign in the Kasner exponents, implies $\fbp - \sum H_i \to 0$ in the initial data at $t \ra -\infty$.  A simple example is one  in which the initial  configuration at large negative times describes $d=3$ contracting dimensions with scale factor $a_1 \sim (-t)^{1/\sqrt{d+n}}$, and $n=6$ expanding dimensions with scale factor $a_2 \sim (-t)^{-1/\sqrt{d+n}}$ (which, together with $\fb \sim - \ln(-t)$, is an exact solution of the vacuum low-energy equations \cite{7,8,9}). By numerically integrating as in the previous example we can check that the solution implements a smooth transition to the duality-related  post-big bang configuration described, at large positive times, by $a_1 \sim (t)^{-1/\sqrt{d+n}}$, $a_2 \sim (t)^{1/\sqrt{d+n}}$, $\fb \sim - \ln(t)$. The overall time evolution of $\dot \phi$ (normalised to $\fbp$) is illustrated in Fig. \ref{f7a}. Initially, the dilaton's evolution is comparable to that of the background geometry,  but stops in the final asymptotic regime after the bounce.

%%%%%%%%%%%%%%%%%%%%%%%%%%%%%%%

\section{Summary and outlook}
\label{sec4}

In this short note we have made a first exploratory study of a large class of string cosmology scenarios which, thanks to all-order $\ap$ corrections, realize the old pre-big bang idea \cite{5, 7} of smoothly joining two duality-related solutions which would be otherwise separated by a curvature singularity.

In our opinion this result removes the main conceptual shortcoming of the PBB scenario, its lack of a consistent description of the high-curvature phase, opening the way to the computation of physical effects originating from that phase.

On the technical side this modest breakthrough was made possible by two crucial ingredients: $i)$ the possibility, following Hohm and Zwiebach \cite{1}, to represent in a compact and handy way, the effect of $\ap$ corrections to all orders through a single duality-invariant function of the Hubble parameter (or parameters in the anisotropic case), and $ii)$ the realisation that the apparent obstacle to achieving a smooth bounce disappears if one looks at some kind of Hamiltonian corresponding to a Legendre transform of the Hohm-Zwiebach Lagrangian and notices that, under mild restrictions, it predicts a ``clockwise"  rather than the apparently more natural ``anti-clockwise" bounce that one had been looking for all the time\footnote{We stress again that the clockwise trajectory refers here to a  geometry undergoing a transition from an initial expanding to a final expanding state. The opposite is true, of course, for a trajectory connecting an initial contracting to a final contracting configuration: its  path cannot be confined to the half plane $H < 0$, and its shape resembles an  ``inverted" heart.}. For this to happen, the Hamiltonian of the system can be a quite simple analytic function. However, when inverting the Legendre transform, it implies the presence of branch point singularities in the Hohm-Zwiebach Lagrangian\footnote{The possible link between non-singular cosmological solutions and multi-valued functions has been discussed, in the context of ``mimetic gravity" \cite{33},   in \cite{34,35}. We thank Jerome Quintin for bringing these papers to our attention.}.

In physical terms this means that the regular bounce is achieved {\it after} the initial (string-frame) accelerated expansion reaches a maximal expansion rate and turns into a (string-frame) contraction before making the real bounce when the contraction rate hits its maximal (string-size) value (see Figs. \ref{f4}, \ref{f5}, \ref{f6}).

At lowest order in the loop expansion, the existence of an exact symmetry \cite{Sen} makes the cosmology after the bounce a mirror image of the one before. But, as noticed long ago, the mirror evolution takes inevitably the solution into the strong coupling regime in which loop and non-perturbative effects in the string coupling become important and break the symmetry between the pre and post-bounce phase. Loop effects will bring particle and entropy production, inhomogeneities, and non perturbative potentials possibly stabilising the dilaton and other moduli. The big question is: will this produce a post-big bang scenario consistent with observations? Having a solid, zero order approximation makes this, at least,  a well-posed question.

Let us however conclude with a word of warning: In order to achieve the non-singular bounce we had to make some assumptions on the (Legendre transform of the) Hohm-Zwiebach function. String theory could be nasty enough to prevent these simple properties to be realized. Also, it is not clear that any choice for the  Hohm-Zwiebach function is compatible with the cosmological reduction of a general covariant action, a  weaker requirement than asking it to follow from some particular string theory.
We hope that the results we presented here will encourage people to address these difficult, but possibly very rewarding, questions.

%%%%%%%%%%%%%
%%%%%%%%%%%%%

\appendix
\section{Mathematical obstructions to an expanding anti-clockwise bounce}
\label{app}

In this Appendix we give a ``physicist proof" that an always expanding ``anti-clockwise bounce" of the type depicted in Fig. \ref{f1} (i.e.  a bounce staying all the time in the upper half plane of that picture)  is incompatible with the  duality-invariant cosmological equations (\ref{1}).  Our proof is by contradiction and goes as follows\footnote{Related arguments reaching similar conclusions for single-valued functions $f(H)$ have also been given in \cite{4}.}.
 
 In the far past $\fbp >0$. In order for $ \fbp$ to turn eventually negative there must be a time interval during which  $\fbpp < 0$. However, if $H$ remains all the time positive, as we assumed, the third of Eqs. {(\ref{1}) implies  $f > 0$ during that same time interval. 

 %%%%%%%%%%%%%%%%%%%%%%%%
 \begin{figure}[t]
\centering
\includegraphics[width=8cm]{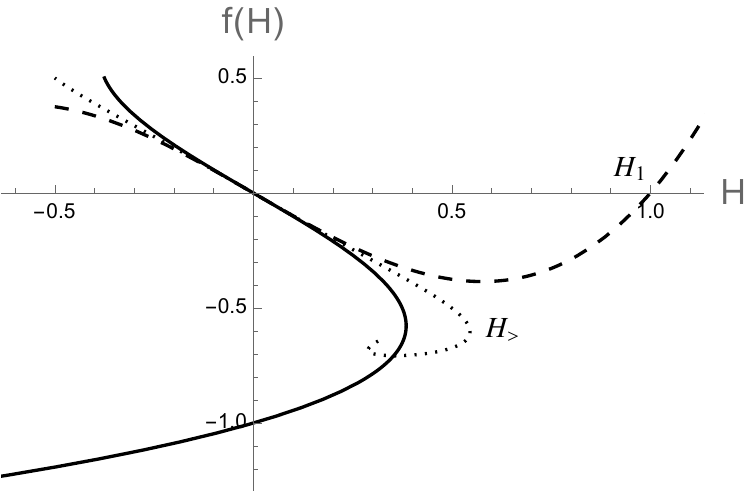}
\caption{Three possible plots of $f$ versus $H$. The dashed curve is the case of a single-valued $f(H)$ staying at positive values of $H$ during the time evolution. The solid curve corresponds to a single-valued $H(f)$ and leads to the regular bounce discussed in the main text. Finally, the dotted curve is a more complicated curve which avoids negative values of $H$ but gives a curvature singularity at a finite $t$.}
\label{f8}
\end{figure} 
%%%%%%%%%%%%%%%%%%%%%

 Let us first assume $f(H)$ to be a regular, single-valued function. This possibility is illustrated by the dashed curve of Fig. \ref{f8}. In this case,
 since $f$ is negative at very small $H$, $f'$ must turn positive at some point and $f$ itself must go through zero at some positive value of $H$. Let's denote by $H_1 > 0$ the first zero of $f$ (i.e. $f(H_1)=0$ and $f(H) < 0$ for $0 < H < H_1$, see the dashed curve in Fig. \ref{f8}).  
  Consider now the quantity $-g(H) \equiv (F - H f)$ appearing in the first of Eqs. (\ref{1}). The quantity $-g$} is positive at small $H$ and its value at $H = H_1$ is clearly $F(H_1)$. Since $F(0) =0$ we  can compute it as:
 \beq
-g(H_1) = F(H_1) = \int_0^{H_1}  f(\tilde{H}) d \tilde{H} <0\,,
\label{A1}
\eeq
 since the integrand is always negative. But then there should be some positive $H_2 < H_1$ at which $g(H_2) =0$ and thus, by the first of Eqs. (\ref{1}), $\fbp = 0$ at the time $t_2$ at which $H= H_2$. Finally, since $\fbp$ is initially positive, there must be a $t_3 < t_2< t_1$ at which $\fbpp =0$, in contradiction with having $\fbpp <0$ as long as $H >0$ and $\fbp <0$ (which is the case up to time $t_3$). We have thus arrived at a contradiction (in agreement with the results of \cite{4}). 

 Consider now the case in which $f(H)$ is not single valued.
% \footnote{The possible link between non-singular cosmological solutions and multi-valued functions has been discussed (but in a different context) also in \cite{33,34,35}. We thank Jerome Quintin for pointing out these papers to our attention.} 
 This looks actually necessary if we want the evolution to go through negative values of $H$. Furthermore, such an $f(H)$ will typically exhibit branch points. This possibility is illustrated by the solid and dotted curves in Fig. (\ref{f8}) with the branch point occurring where the curve $f(H)$ turns around at $H=H_>$. At that point $f' \to \infty$ and $f$ changes branch at the corresponding time $t_>$. For $t >t_>$, $f$ keeps decreasing in time but so does $H$, so that $f'(H) \equiv df/dH > 0$ for $t >t_>$. We have thus achieved the desired change of sign in $f'$ in a different way w.r.t. the dashed curve. In our regular bounce scenarios  (solid curve in Fig. (\ref{f8})) one eventually goes through $H=0$ into a contracting phase and everything works fine until $H$ reaches its minimum negative value $-H_<$ at the regular bounce.

One can ask whether, instead, one can avoid crossing the $H =0$ axis, the possibility we claim to be able to exclude. That would require $f'(H)$ to change sign again before reaching the $H=0$ axis (dotted curve in Fig. \ref{f8}) but  this would imply, by the second of Eqs. (\ref{1}), a divergent $\dot{H}$, thus a curvature singularity, at the time at which $f'(H)=0$. 

Thus, without claiming to have given a rigorous mathematical proof, we seem to have very strong arguments in favour of excluding a smooth transition 
from accelerated expansion to decelerated expansion involving only the upper half of the plane depicted in Fig. \ref{f1} for any string cosmology model which is duality-invariant at all orders in $\ap$.

%%%%%%%%%%%%%%%%%%%%%%%%%%%%%%%%

\acknowledgments

MG  is supported in part by INFN under the program TAsP: {\it ``Theoretical Astroparticle Physics"}, and by the research grant number 2017W4HA7S {\it ``NAT-NET: Neutrino and Astroparticle Theory Network"}, under the program PRIN 2017 funded by the Italian Ministero dell'Universit\`a e della Ricerca (MUR). We wish to thank the hospitality and financial support of the Dipartimento di Fisica and Sezione INFN di Pisa, and Pietro Conzinu, Giuseppe Fanizza and Giovanni Marozzi for useful discussions.
Finally, we are grateful to Robert Brandenberger, Krzysztof A. Meissner, Jerome Quintin, Peng Wang, Houwen Wu, Haitang Yang, Shuxuan Ying  and Barton Zwiebach for their interesting comments and suggestions.

\end{document}